# Teaching the Foundations of Data Science: An Interdisciplinary Approach

*Research-in-Progress*

**Daniel Adomako Asamoah**

Department of Information Systems & Supply Chain Management
Raj Soin College of Business
Wright State University

**Derek Doran**

Department of Computer Science & Engineering
Kno.e.sis Research Center
Wright State University

**Shu Schiller**

Department of Information Systems & Supply Chain Management
Raj Soin College of Business
Wright State University

## Introduction

The paucity of human resource skills to acquire, effectively analyze, and extract valuable insights from data has overtaken data acquisition as the principal challenge facing data-inspired analysis in organizations (Dhar 2012). This scarcity motivates the need for training well-qualified data scientists able to generate insights from large amounts of complex data hoarded by organizations (Davenport and Patil 2012). In 2011, the McKinsey Global Institute reported a shortage of between 140,000 and 190,000 workers competent in data analytics (Manyika, Chui, Brown, Bughin, Dobbs, and Roxburgh 2011).

The analytics industry has established some lead in producing "data-based" decision support experts, most of whom are trained in-house in a specific domain with technologies from a specific vendor. Such training, however, is unlikely to produce an overarching impact in both research and practice. The industry needs a well-structured approach for talent development that imparts the utilization of computational and statistical methods and an ability to apply these methods to a specific domain. Academic "data science" programs and courses attempt to address this problem, but most are delivered with emphasis on a specific perspective, likely narrowed down by the principal domain of the instructor's academic discipline (Cleveland 2001).

An interdisciplinary aspect of data science courses and curriculums, where domain-specific and technical skills are imparted to students, is essential. This need is based on two very broad perspectives on the role of a "data scientist" that exist: a "practical perspective" where data scientists are seen as business analysts that harvest actionable insights from scores of high-dimensional, possibly heterogeneous data; and a "technical perspective" where data scientists are seen as experts in advanced computational tools, data mining algorithms, statistical analysis, and machine learning. Successful data scientists must span both perspectives in practice, as they use business domain expertise, advanced computing, mathematical, and statistical methods to find knowledge that they themselves synthesize and report on (Conway 2010). We hypothesize that a course to introduce data science, developed and delivered in an interdisciplinary fashion between Management Information Systems (MIS) and Computer Science (CS), can train students with skills, knowledge, and ability that are a marriage of these practical and technical perspectives and may thus serve as a strong foundation for further education and training on all aspects of data science.

This paper shares our perspectives and experiences in developing and offering an interdisciplinary course in data science, targeting students with a semi-developed computing and mathematical background that matches the typical prerequisites for MIS or CS junior courses. The course imparts the minimum skill set necessary to ingest, analyze, and make predictions over datasets in practice and the background necessary to pursue an advanced data science certificate or a graduate program. In its first offering, course participants were a cross-disciplinary group of students from the colleges of business, liberal arts, and engineering and computer science. The instructors, one MIS professor and the other, a CS professor, leveraged their individual expertise and experiences from their respective business and engineering disciplines to create an interdisciplinary synergy. Although students belonged to different majors, the assignments, projects, and class discussions were carried out collaboratively.

*Teaching Data Science: An Interdisciplinary Approach*This paper is based on a research-in-progress, and is organized as follows. First, we present a literature review of different frameworks and motivations for the training of an interdisciplinary data scientist. We then present our argument of how existing approaches can be enhanced with the integration of an interdisciplinary approach. Next, we share our experiences of applying the interdisciplinary approach in the development and teaching of a data science course, including specific course objectives, course format and structure, curriculum design, and our implementation strategy. This is followed by an analysis of the assessment of learning outcomes. Finally, we conclude with a discussion on the challenges and lessons learned in our journey.

## Literature Review

Data science is defined as the use of quantitative and qualitative methods and strategies on data to solve relevant problems to predict outcomes (Waller and Fawcett 2013). The methods encompass statistics, CS, and strategies for understanding and explaining phenomena in a particular domain of interest. Therefore, by extension, a data scientist is one who has skills in programming, statistics, and has developed domain expertise in a relevant area. They must further draw on such skills to effectively acquire, manage, analyze, and generate insights from data. These requirements lend themselves to an interdisciplinary approach to teaching the subject. However, the question still remains as to how institutions can deliver such an interdisciplinary course or program to students who are traditionally comfortable with learning within the context of a single academic discipline and program.

To meet the rising demand for data experts in the industry, data science has been delivered under different academic umbrellas and in multiple formats. Some institutions offer comprehensive degree programs in data science (Anderson et al. 2014) whereas others offer only individual courses as part of a broader academic program (Kolluru et al. 2012). In some institutions, the data science course and/or program is categorized as belonging to only certain departments. For instance, Chiang et al. (2012) proposes that data science courses should be sidelined primarily to CS departments owing to its technical nature while others suggest a business focus in the education of big data and advanced business intelligence (Gupta et al. 2015). Regardless of the arguments on the positioning of data science education, multiple calls have proposed teaching data science as an interdisciplinary course. Based on Davis (1995), we define interdisciplinary courses as those that are created as a result of scholars from at least two disciplines coherently combining their perspectives to achieve a common pedagogic goal. It is imperative that an interdisciplinary data science course be designed holistically, rather than stringing together of individual courses or topics (Ivanitskaya et al. 2002). Hence, both the Association to Advance Collegiate Schools of Business (AACSB) and the Accreditation Board for Engineering and Technology (ABET) recognize the interdisciplinary nature of education as a key dimension in their accreditation procedures (Borrego et al. 2007; AACSB International 2005).

Previous literature has conceptualized and categorized multiple frameworks for interdisciplinary pedagogy. For instance, Lattuca et al. (2004) created a topology of four forms of interdisciplinarity. They describe the first, informed disciplinarity as being situated in a single disciplinary unit even though an instructor could leverage content from other disciplines to explain concepts. They describe *transdisicplinarity* as one that amalgamates theories and methods from other disciplines into one and applies them to several other disciplines. In so doing, the connection to resident sources and disciplines from which the theories and methods originated are erased. Next is synthetic interdisciplinarity, in which even though theories and concepts are combined from other disciplines, the originating disciplines effectively maintain their identity by retaining distinct content, boundaries, theories and methods. Lastly, they presented conceptual interdisciplinarity, in which there is a loose reference to other disciplines without any strong emphasis on any particular one. In our course, we employed synthetic interdisciplinarity given the distinctive nature of the individual fields of statistics, computer science and the multiple substantive domain areas we focused on.

Way and Whidden (2014) introduced a loosely-coupled interdisciplinary approach in which two distinct courses are delivered in parallel to different groups, yet with a "merge point," where the two classes come together at specific periods for combined learning sessions. This approach was implemented in a crosscutting course on game design in. It is contrasted against a tightly coupled approach where two computing topics or courses are merged for the entire duration of the course as in (Counsell 2003). Despite the challenges of planning and delivering an interdisciplinary course, studies have found that

**2**   *Pre-ICIS SIGDSA Workshop, Fort Worth, Texas 2015*

such an environment helps leaners gain knowledge and skills that allow them to inculcate cross-disciplinary theories and perspectives (Lattuca et al. 2004, So and Brush 2008). They also foster higher order thinking and improve critical thinking skills.

Cassel (2011), who advocates for interdisciplinarity in computing curriculums, argues for the merging of silos in disciplines regardless of the varied goals and motivations of the disciplines. A course in data science is particularly amenable for such merging. Courses in data science offered at several institutions (Hardin et al. 2014) exhibit different degrees of interdisciplinarity, some of which take place at the academic program level (e.g., creating an interdisciplinary data science graduate certificate or Master's degree program). In this paper, we introduce interdisciplinary data science curriculum at the course level – an "Introduction to Data Science" class for upper-level undergraduate and graduate level students in multiple colleges. The course is interdisciplinary along many facets: (i) the instructors originate from different academic backgrounds (one MIS and the other CS) and utilize their research and teaching experience collaboratively in course development and delivery; (ii) its content consists of concepts that span business, computing, mathematical, and interpretation and reporting skills; (iii) students who took the class majored in multiple academic disciplines.

## The "Introduction to Data Science" Course

We cooperatively developed and delivered an introduction to data science course at Wright State University (WSU) in the summer of 2015. The course objectives emerged from the advances of data science curriculum development, inputs from corporate advisory boards, and our personal experience in data analytics research (Asamoah and Sharda 2013; Doran and Gokhale 2008; Lin et al. 2014; Sharda et al. 2013) and practice (Doran et al. 2012). The course objectives are categorized into facets that often describe a data scientist (Conway 2010):

- <u>Math & Statistics Knowledge:</u> Students should develop a toolkit of techniques for running summary, classification, and predictive analytics over (un)structured data; be able to construct basic models of data for analysis; gain experience with pattern recognition and knowledge discovery techniques to establish data relationships; demonstrate data analysis techniques and simulations of phenomena of interest in engineering in science.
- <u>Hacking skills:</u> Students should become familiar with modern open-source programming languages to run data analytics; learn high level tools to visualize data to gain insight into data relationships; understand the process and tools supporting essential data analytics tasks, including extraction, cleaning, classification or prediction, and interpretation.
- <u>Substantive/business expertise</u>: Students should appreciate the potential analytics approach to solving business problems in today's data rich environment; understand how to interpret analysis results and convey the findings to a layman; and apply science and business principles to analyze and interpret data using analytical and computer-based techniques and applications.

We decided that the course objectives would be best achieved by a series of hands-on laboratory sessions (eight in total) and a semester long group project, rather than individualized written and programming assignments. Laboratory sessions trained students on the use of data science tools in a collaborative environment and offered practice on the reporting and interpretation of results on real data. Every laboratory assignment included the manipulation of real data using R and/or Python. To promote interdisciplinary learning, students were encouraged to work on laboratory assignments in small groups. In addition to the lab sessions, students needed to complete a group project that supported the objectives of substantive expertise by engaging the students with defining and exploring in-depth a dataset within the business domain it had emerged from. Project teams were formed with mixed-majors students, which consisted of at least two students from different academic. In the following sub-sections, we offer details about the structure of the course, the technologies used in the lab sessions, and the semester long project.

### *Course Structure*

With MIS and CS majors as the target audience, we investigated the course programs of each major to search for intersecting skills developed within them. In either major, we found that WSU students are not required to take courses in probability theory, multivariable calculus, or in a statistics course at the calculus level (e.g., a statistic courses that discusses concepts such as a distribution or density function, maximum likelihood estimation, and common discrete and continuous distributions). Courses on data collection, ingestion, and summarization using computing platforms are also not required. Finally, MIS students need only take a small number of programming courses; however, they are well-placed in reviewing analytics problems from a business perspective. We postulate that these observations match the intersecting background of MIS and CS students in a majority of undergraduate teaching institutions. Based on these observations, we defined a course structure (shown in Figure 1) that builds students' technical skills from the ground up. It begins with learning to use computational tools to manipulate and run scripts over data; then to use essential mathematical and statistics concepts before discussing the predictive analytics methods that are built upon them. Laboratories with real data are used in each part of the course to practice the discussed concepts.

| Course Components | Practice Domain/Cases | |
|---|---|---|
| Part 3: Predictive Analytics<br>Linear Regression, Logistic Regression, SVMs, k-means Clustering | Stock market returns; commuter cars; supermarket shopping | Semester Long project: Wine quality; healthcare |
| Part 2: Math & Statistics Foundations<br>Probabilistic Modeling and Statistical Analysis | Gambling simulations; physical fitness data; sporting events | |
| Part 1: Data Hacking<br>Computational Tools; Data Wrangling; Visualization | Movie reviews; city crime reports | |

Substantive expertise developed by studying and interpreting real datasets

**Figure 1: Structure of "Introduction to Data Science" course. Practice in the analysis of real life datasets are carried out through interactive labs and a course project.**

Table 1 lists the specific topics covered in parts 1, 2, and 3 of the course. The Data Hacking part emphasized the use of R and Python as data analysis tools, and their capabilities for data preprocessing and visualization. The Math & Statistics Foundations part shored up students foundation in probability theory and in statistical methods such as hypothesis testing, t-tests, and ANOVA. The Predictive Analytics part of the course discussed the theoretical and practical aspects of running and interpreting models for predictive analytics. Approximately 40% of the course was devoted to data hacking, 20% to math and statistics foundations, and 40% to predictive analytics. The larger percentage of time spent on data hacking ensured that students had the opportunity to develop their skills of manipulating data with R and Python. Predictive analytics was also given more time because the material covered how learning algorithms operated theoretically as well as how to run the algorithms in R and interpret their output. Students developed the 'soft kills' substantive experts exhibit by evaluating datasets spanning business contexts such as movie reviews, crime reports, gambling scenarios, and supermarket shopping behaviors.

## *Course Project*

The course project required students to define hypotheses within a domain that they are interested in, collect data from that domain, and analyze the data to test their hypotheses. Emphasis was placed on 'story telling through data' and on applying data hacking and predictive analytics techniques learned from lectures. We enforced mixed majors of CS and MIS students in the same group to promote interdisciplinary thinking. We requested the students to collect their own datasets, process and analyze it. To streamline the project process and save time on data collection, we also provided multiple data sets to students as an option. Specifically, students could choose the data on home healthcare services or the quality of European wines. No matter the dataset chosen, students still needed to perform data preparation, manipulation, and analytics. Students presented a project proposal at the beginning of the course and a final presentation during the last week of the semester. Both presentations were evaluated using the following criteria:

- Did the presentation have a clear opening statement, maintain focus, and clearly express the hypotheses to be explored? Was the presentation well organized?
- Did the group collected the necessary background knowledge to formulate reasonable hypothesis? Did they demonstrate that they have accrued domain-specific knowledge? In the final presentation, did the group demonstrate a comprehensive understanding of the domain the data has come from?
- Were the hypotheses sufficiently impactful or challenging? In the final presentation, was the depth of the analysis appropriate? Did the group use effective visuals and tell a compelling story? Would a manager or decision maker decide to take action based on the results of their analysis?

| **Part 1 – Data Hacking** |
|---|
| ***Introduction to R***: Introduces R as a tool to run statistical analysis and as a full-fledged scripting language for developing complex data analysis workflows. Discussed R data structures (vectors, data frames, lists, and matrices), conditional and control flow structures, logical indexing into data structures, and elementary scatter plotting. |
| ***Introduction to Python***: Introduced Python, its data structures, string processing routines and packages for collecting, cleaning and preprocessing data from online APIs. |
| ***Data collection and wrangling***: Presented the form of data interchange formats (XML and JSON), and their manipulation in Python and R. Discussed the data heterogeneity problem, where a data scientist must synthesize data from independent sources into a single set for analysis, and data merging and reshaping in R. |
| ***Data visualization***: Discussed best practices in visualization and how to make it appealing. The ggplot2 package for data visualizations and for plotting geo-spatial data in R was emphasized. |
| **Part 2 – Mathematics and Statistics Foundations** |
| ***Probability Theory***: Discussed elementary concepts of discrete probability theory, including basic combinatorics, notions of sample and event spaces, the axioms of probability, and Bayes' theorem. |
| ***Probabilistic Modeling***: Discussed concepts of random variables and discrete probability distributions. The notion of expectation, utility, and risk is used in examples from decision science. |
| ***Statistical Analysis***: Discussed sample statistics, sampling numeric and categorical variables, and hypothesis testing involving one and two group means t-tests, one-way ANOVA, factorial ANOVA, and repeated measures ANOVA. |
| **Part 3 – Predictive Analytics** |
| ***Supervised Learning and Linear Models***: Discussed concepts of building a machine learning algorithm, which includes creating a functional form (i.e. hypothesis) that relates data features to a response variable for prediction. Discussed the output of a linear regression ran in R, including tests for a relationship between a factor and a response variable, and understanding what an $R^2$ value is and how it is derived. |
| ***Classification and Logistic Regression***: Contrasted regression (predicting a real number output) and classification (predicting the value of a discrete response variable). Discussed how logistic regression may be seen as dividing a feature space into regions by a decision boundary, and building and testing the accuracy of a logistic regression classifier in R. |
| ***Support Vector Machines (SVMs)***: Motivated a classification algorithm that finds non-linear classification boundaries in a principled way. Discussed the use of kernels to project data into higher dimensional spaces and how to create and run SVMs in R. |
| ***k-means Clustering***: Introduced the notion of unsupervised learning. Guidelines for assessing the 'quality' of a clustering result were also presented. Discussed how to run k-means clustering in R. |

**Table 1: Topics Covered in Different Phases of the Course**

## Assessment of Learning

We assessed the quality of the course through an end-of-course survey involving 14 registered students, among which 4 were MIS major and 9 were CS major. Table 2 summarizes the demographics of the students. We adopted and adapted items from Ducoffe (2006) and So and Brush (2008), who recommended evaluating the perceptions of students enrolled in the course. Given that the collaborative aspect of our course was emphasized during its design and development, the survey includes additional items related to interdisciplinarity, namely 1) the level of achieved collaboration created among learners; and 2) the level of satisfaction among learners based on predetermined learning goals outlined.

| Gender | | | Age | | | Ethnicity | | |
|---|---|---|---|---|---|---|---|---|
| Male | 10 | 71% | Less than 20 years old | 0 | 0% | African-American | 1 | 7% |
| Female | 4 | 29% | 20 to 24 years | 7 | 50% | Asian / Pacific Islander | 1 | 7% |
| Total | 14 | 100% | 25 to 29 years | 2 | 14% | Caucasian | 10 | 71% |
| | | | 30 years and above | 5 | 36% | Latino | 1 | 7% |
| | | | Total | 14 | 100% | Other | 1 | 7% |
| | | | | | | Prefer not to disclose | 0 | 0% |
| | | | | | | Total | 14 | 100% |
| **Degree Pursuing** | | | **Year in College** | | | **Major** | | |
| Bachelors | 13 | 93% | Freshman | 0 | 0% | MIS | 4 | 29% |
| Masters | 1 | 7% | Sophomore | 1 | 8% | CS | 9 | 64% |
| PhD | 0 | 0% | Junior | 1 | 8% | Statistics | 0 | 0% |
| Other | 0 | 0% | Senior | 10 | 77% | Other | 1 | 7% |
| Total | 14 | 100% | Other | 1 | 8% | Total | 14 | 100% |
| | | | Total | 13 | 100% | | | |

**Table 2: Demographics**

| Item (adapted from So and Brush (2008), Ducoffe et al. (2006)) | Mean | S.D. | Min | Max |
|---|---|---|---|---|
| Satisfaction | 5.23 | 1.60 | 1 | 7 |
| Collaborative Learning | 5.04 | 1.55 | 1 | 7 |
| Collaborative (Team) Teaching | 5.54 | 1.47 | 1 | 7 |
| Interdisciplinary Nature | 5.13 | 1.81 | 1 | 7 |
| Overall Attitude | 5.43 | 1.75 | 1 | 7 |
| Comparison to other types of courses | 5.20 | 1.81 | 1 | 7 |

**Table 3: Assessment of Student Learning Experience**

| Item | Mean | S.D. | Min | Max |
|---|---|---|---|---|
| Identify the advantages and disadvantages of different technical methods to analyze a given dataset | 4.77 | 1.79 | 1 | 7 |
| Use modern open-source tools to collect, process, store, and analyze high dimension data | 5.23 | 1.54 | 2 | 7 |
| Apply science and business principles to analyze and interpret data, using analytic and computer-based techniques | 5.15 | 1.68 | 1 | 7 |
| Apply data mining techniques to generate interesting business insights in all volumes of data | 4.85 | 1.77 | 2 | 7 |
| Effectively interpret and communicate my ideas through written and oral reports | 5.23 | 1.64 | 1 | 7 |
| Understand how to appropriately test hypotheses and interpret predictive models that are applied to data sets. | 4.92 | 1.61 | 1 | 7 |

**Table 4: Assessment of Student Learning Outcomes**

Students' learning experience was evaluated from six aspects, each of which consists of multiple measurement items (7 Likert scale) adapted from (Ducoffe 2006) and (So and Brush 2008). In general, students felt very satisfied with the course, its interdisciplinary nature, the co-learning and co-teaching model, and developed a positive attitude toward the learning experience (Table 3). Student's particularly found the course to improve their attitude about data science as a possible career choice, and found the team teaching methods to positively enrich their learning. Students also responded positively, albeit with a weaker response, to the collaborative learning of the course. From our own observations on students in the class, we postulate that some student groups may have had challenges interacting with students who came from a different major and hence had a different skill set and background. In fact, this was likely the first experience of working in an interdisciplinary team for most students. Future offerings of this course will therefore also discuss strategies for working in an interdisciplinary fashion.

Students' individual learning outcomes were measured on six items that are based on the learning goals of the course. The self-reported responses suggest that students were able to master knowledge and skills on various aspects of data science (Table 4). This course format was particularly capable of conveying skills on the use of open-source tools to run data analytics, to apply business principles to study and interpret data, and to effectively communicate their findings through written and oral reports. Outcomes on running data mining and predictive analytics methods over the data are lower, perhaps due to the fact that these concepts are emphasized at the end of the course whereas computational tools and data interpretation are emphasized throughout. Future offerings of the course may devote more time to these aspects in order to improve student learning outcomes.

Ultimately, the assessment on learning experience and outcomes showed satisfactory results on the co-teaching, co-learning approach. Although the class size was small with 14 students, the course provided some valuable empirical data for the effectiveness of an interdisciplinary data science course and built a foundation for future improvements. Furthermore, students enjoyed the co-teaching style and greatly improved their knowledge and skill sets on data science through hands-on practice with lab sessions and course project. One limitation of this paper is the small sample size used to assess student satisfaction. We intend to gather more assessment data in future offerings of the course so as to be able to make more generalizable evaluations about our approach to teaching an interdisciplinary course. Our experiences and approach, as outlines in this paper, can be extended to other fields of study where interdisciplinarity is paramount for student satisfaction and success.

## Conclusion and Future Work

This paper introduced our experience in designing, developing, and delivering an interdisciplinary Introduction to Data Science course. Due to the nature of data science, the course had a focus on both practical and technical perspectives, integrating the expertise from both MIS and CS and was co-taught by two instructors in the two domains with the synthetic interdisciplinarity approach (Lattuca et al. 2004). The course covered topics from data manipulation, management, visualization, and preprocessing using modern tools, foundational knowledge in probability and statistics, and the theoretical and practical aspects of predictive analytics algorithms. A post-course survey revealed that students were satisfied with the content of the course, the material learned, and responded positively to its interdisciplinary nature.

Some intriguing questions will guide us in our future studies. For instance, we are interested in potential differences between MIS and CS students. Were MIS students successful at the same level with CS students, given different backgrounds? How can the course be designed to strengthen their existing skills and to grow their new skills respectively? What improvements can be made to support their "team" learning with domain knowledge from different subject fields? In addition, we hope to identify future opportunities to refine our pedagogy and to examine and generalize the impact of different domains of study on student performance in interdisciplinary courses beyond the subject of data science, with the potential value to enrich other fields of study where interdisciplinarity plays a central role.

## References

Anderson, P., Bowring, J., McCauley, R., Pothering, G., and Starr, C. 2014. "An Undergraduate Degree in Data Science: Curriculum and a Decade of Implementation Experience," in *Proceedings of the 45th*

ACM technical symposium on Computer science Education, ACM, pp. 145–150.

Asamoah, D., and Sharda, R. 2013. "Predicting Injury Recovery Time in Collegiate Football: A Data Mining Approach," in *Eleventh Annual Big XII+ MIS Research Symposium*.

Borrego, M., Newswander, L., and McNair, L. D. 2007. "Special session - Applying theories of interdisciplinary collaboration in research and teaching practice," in *Proceedings - Frontiers in Education Conference, FIE*.

Cassel, L. N. 2011. "Interdisciplinary Computing is the Answer: Now, What was the Question?," *ACM Inroads* (2:1), ACM, pp. 4–6.

Chiang, R. H. L., Goes, P., and Stohr, E. A. 2012. "Business Intelligence and Analytics Education, and Program Development: A Unique Opportunity for the Information Systems Discipline," *ACM Trans. Manage. Inf. Syst.* (3:3), pp. 12:1–12:13 (available at http://doi.acm.org/10.1145/2361256.2361257).

Cleveland, W. S. 2001. "Data Science: An Action Plan for Expanding the Technical Areas of the Field of Statistics," *International statistical review* (69:1), Wiley Online Library, pp. 21–26.

Conway, D. 2010. "The Data Science Venn Diagram," *dataists* (available at http://www.dataists.com/2010/09/the-data-science-venn-diagram/).

Counsell, D. 2003. "A Review of Bioinformatics Education in the UK," *Briefings in Bioinformatics* (4:1), pp. 7–21.

Davenport, T. H., and Patil, D. J. 2012. "Data scientist: The Sexiest Job of the 21st Century.," *Harvard business review* (90:10).

Davis, J. R. 1995. *Interdisciplinary Courses and Team Teaching: New Arrangements for Learning*, American Council on Education and the Oryx Press Phoenix, AZ.

Dhar, V. 2012. "Data Science and Prediction," *Communications of the ACM* (56:12), pp. 64–73 (available at http://www.ssrn.com/abstract=2086734).

Doran, D., and Gokhale, S. S. 2008. "Discovering New Trends in Web Robot Traffic through Functional Classification," in *Network Computing and Applications, 2008. NCA'08. Seventh IEEE International Symposium on*, IEEE, pp. 275–278.

Doran, D., Mendiratta, V., Phadke, C., and Uzunalioglu, H. 2012. "The Importance of Outlier Relationships in Mobile Call Graphs," in *Machine Learning and Applications (ICMLA), 2012 11th International Conference on* (Vol. 2), IEEE, pp. 24–29.

Ducoffe, S. J. S. 2006. "Interdisciplinary, Team-Taught, Undergraduate Business Courses: The Impact of Integration," *Journal of Management Education*, pp. 276–294.

Gupta, B., Goul, M., and Dinter, B. 2015. "Business Intelligence and Big Data in Higher Education: Status of a Multi-Year Model Curriculum Development Effort for Business School Undergraduates, MS Graduates, and MBAs," *Communications of the Association for Information Systems* (36:1), p. 23.

Hardin, J., Hoerl, R., Horton, N. J., and Nolan, D. 2014. "Data Science in the Statistics Curricula: Preparing Students to 'Think with Data,'" *arXiv preprint arXiv:1410.3127*.

International, A. 2005. *Eligibility Procedures and Accreditation Standards for Business Accreditation*, AACSB International.

Ivanitskaya, L., Clark, D., Montgomery, G., and Primeau, R. 2002. "Interdisciplinary Learning : Process and Outcomes," *Innovative Higher Education* (27:2), pp. 95–111.

James Manyika, Michael Chui, Brad Brown, Jacques Bughin, Richard Dobbs, Charles Roxburgh, A. H. B. 2011. "Big Data: The Next Frontier for Innovation, Competition, and Productivity," *McKinsey Global Institute*, p. 156 (available at http://scholar.google.com/scholar.bib?q=info:kkCtazs1Q6wJ:scholar.google.com/&output=citation&hl=en&as_sdt=0,47&ct=citation&cd=0).

Kolluru, S., Roesch, D. M., and de la Fuente, A. A. 2012. "A Multi-Instructor, Team-Based, Active-Learning Exercise to Integrate Basic and Clinical Sciences Content," *American journal of pharmaceutical education* (76:2), American Association of Colleges of Pharmacy.

Lattuca, L. R., Voigt, L. J., and Fath, K. Q. 2004. "Does Interdisciplinarity Promote Learning? Theoretical Support and Researchable Questions," *The Review of Higher Education* (28:1), The Johns Hopkins University Press, pp. 23–48.

Lin, L., Dagnino, A., Doran, D., and Gokhale, S. 2014. "Data Analytics for Power Utility Storm Planning," in *Proc. of International Conference on Knowledge Discovery and Information Retrieval*, Rome, Italy: SCITEPRESS Digital Proceedings.

Sharda, R., Asamoah, D. A., and Ponna, N. 2013. "Research and Pedagogy in Business Analytics: Opportunities and Illustrative Examples," *Journal of Computing and Information Technology* (21:3), SRCE-Sveučilišni računski centar, pp. 171–183.

So, H.-J., and Brush, T. A. 2008. "Student Perceptions of Collaborative Learning, Social Presence and Satisfaction in a Blended Learning Environment: Relationships and Critical Factors," *Computers & Education* (51:1), Elsevier, pp. 318–336.

Waller, M. A., and Fawcett, S. E. 2013. "Data Science, Predictive Analytics, and Big Data: A Revolution That Will Transform Supply Chain Design and Management," *Journal of Business Logistics* (34:2), pp. 77–84.

Way, T., and Whidden, S. 2014. "A Loosely-Coupled Approach to Interdisciplinary Computer Science Education," in *Proceedings of the International Conference on Frontiers in Education: Computer Science and Computer Engineering (FECS)*, The Steering Committee of The World Congress in Computer Science, Computer Engineering and Applied Computing (WorldComp), p. 1.